\documentclass[letterpaper,compsoc,twoside]{IEEEtran}
\usepackage{fixltx2e} 
\usepackage{cmap} 
\usepackage{ifthen}
\usepackage[T1]{fontenc}
\usepackage[utf8]{inputenc}
\usepackage{amsmath}

\usepackage[font={small,it},labelfont=bf]{caption}
\usepackage{float}

\pdfoutput=1
\usepackage{scipy}
\makeatletter
\def\PY@reset{\let\PY@it=\relax \let\PY@bf=\relax%
    \let\PY@ul=\relax \let\PY@tc=\relax%
    \let\PY@bc=\relax \let\PY@ff=\relax}
\def\PY@tok#1{\csname PY@tok@#1\endcsname}
\def\PY@toks#1+{\ifx\relax#1\empty\else%
    \PY@tok{#1}\expandafter\PY@toks\fi}
\def\PY@do#1{\PY@bc{\PY@tc{\PY@ul{%
    \PY@it{\PY@bf{\PY@ff{#1}}}}}}}
\def\PY#1#2{\PY@reset\PY@toks#1+\relax+\PY@do{#2}}

\expandafter\def\csname PY@tok@gd\endcsname{\def\PY@tc##1{\textcolor[rgb]{0.63,0.00,0.00}{##1}}}
\expandafter\def\csname PY@tok@gu\endcsname{\let\PY@bf=\textbf\def\PY@tc##1{\textcolor[rgb]{0.50,0.00,0.50}{##1}}}
\expandafter\def\csname PY@tok@gt\endcsname{\def\PY@tc##1{\textcolor[rgb]{0.00,0.27,0.87}{##1}}}
\expandafter\def\csname PY@tok@gs\endcsname{\let\PY@bf=\textbf}
\expandafter\def\csname PY@tok@gr\endcsname{\def\PY@tc##1{\textcolor[rgb]{1.00,0.00,0.00}{##1}}}
\expandafter\def\csname PY@tok@cm\endcsname{\let\PY@it=\textit\def\PY@tc##1{\textcolor[rgb]{0.25,0.50,0.56}{##1}}}
\expandafter\def\csname PY@tok@vg\endcsname{\def\PY@tc##1{\textcolor[rgb]{0.73,0.38,0.84}{##1}}}
\expandafter\def\csname PY@tok@m\endcsname{\def\PY@tc##1{\textcolor[rgb]{0.13,0.50,0.31}{##1}}}
\expandafter\def\csname PY@tok@mh\endcsname{\def\PY@tc##1{\textcolor[rgb]{0.13,0.50,0.31}{##1}}}
\expandafter\def\csname PY@tok@cs\endcsname{\def\PY@tc##1{\textcolor[rgb]{0.25,0.50,0.56}{##1}}\def\PY@bc##1{\setlength{\fboxsep}{0pt}\colorbox[rgb]{1.00,0.94,0.94}{\strut ##1}}}
\expandafter\def\csname PY@tok@ge\endcsname{\let\PY@it=\textit}
\expandafter\def\csname PY@tok@vc\endcsname{\def\PY@tc##1{\textcolor[rgb]{0.73,0.38,0.84}{##1}}}
\expandafter\def\csname PY@tok@il\endcsname{\def\PY@tc##1{\textcolor[rgb]{0.13,0.50,0.31}{##1}}}
\expandafter\def\csname PY@tok@go\endcsname{\def\PY@tc##1{\textcolor[rgb]{0.20,0.20,0.20}{##1}}}
\expandafter\def\csname PY@tok@cp\endcsname{\def\PY@tc##1{\textcolor[rgb]{0.00,0.44,0.13}{##1}}}
\expandafter\def\csname PY@tok@gi\endcsname{\def\PY@tc##1{\textcolor[rgb]{0.00,0.63,0.00}{##1}}}
\expandafter\def\csname PY@tok@gh\endcsname{\let\PY@bf=\textbf\def\PY@tc##1{\textcolor[rgb]{0.00,0.00,0.50}{##1}}}
\expandafter\def\csname PY@tok@ni\endcsname{\let\PY@bf=\textbf\def\PY@tc##1{\textcolor[rgb]{0.84,0.33,0.22}{##1}}}
\expandafter\def\csname PY@tok@nl\endcsname{\let\PY@bf=\textbf\def\PY@tc##1{\textcolor[rgb]{0.00,0.13,0.44}{##1}}}
\expandafter\def\csname PY@tok@nn\endcsname{\let\PY@bf=\textbf\def\PY@tc##1{\textcolor[rgb]{0.05,0.52,0.71}{##1}}}
\expandafter\def\csname PY@tok@no\endcsname{\def\PY@tc##1{\textcolor[rgb]{0.38,0.68,0.84}{##1}}}
\expandafter\def\csname PY@tok@na\endcsname{\def\PY@tc##1{\textcolor[rgb]{0.25,0.44,0.63}{##1}}}
\expandafter\def\csname PY@tok@nb\endcsname{\def\PY@tc##1{\textcolor[rgb]{0.00,0.44,0.13}{##1}}}
\expandafter\def\csname PY@tok@nc\endcsname{\let\PY@bf=\textbf\def\PY@tc##1{\textcolor[rgb]{0.05,0.52,0.71}{##1}}}
\expandafter\def\csname PY@tok@nd\endcsname{\let\PY@bf=\textbf\def\PY@tc##1{\textcolor[rgb]{0.33,0.33,0.33}{##1}}}
\expandafter\def\csname PY@tok@ne\endcsname{\def\PY@tc##1{\textcolor[rgb]{0.00,0.44,0.13}{##1}}}
\expandafter\def\csname PY@tok@nf\endcsname{\def\PY@tc##1{\textcolor[rgb]{0.02,0.16,0.49}{##1}}}
\expandafter\def\csname PY@tok@si\endcsname{\let\PY@it=\textit\def\PY@tc##1{\textcolor[rgb]{0.44,0.63,0.82}{##1}}}
\expandafter\def\csname PY@tok@s2\endcsname{\def\PY@tc##1{\textcolor[rgb]{0.25,0.44,0.63}{##1}}}
\expandafter\def\csname PY@tok@vi\endcsname{\def\PY@tc##1{\textcolor[rgb]{0.73,0.38,0.84}{##1}}}
\expandafter\def\csname PY@tok@nt\endcsname{\let\PY@bf=\textbf\def\PY@tc##1{\textcolor[rgb]{0.02,0.16,0.45}{##1}}}
\expandafter\def\csname PY@tok@nv\endcsname{\def\PY@tc##1{\textcolor[rgb]{0.73,0.38,0.84}{##1}}}
\expandafter\def\csname PY@tok@s1\endcsname{\def\PY@tc##1{\textcolor[rgb]{0.25,0.44,0.63}{##1}}}
\expandafter\def\csname PY@tok@gp\endcsname{\let\PY@bf=\textbf\def\PY@tc##1{\textcolor[rgb]{0.78,0.36,0.04}{##1}}}
\expandafter\def\csname PY@tok@sh\endcsname{\def\PY@tc##1{\textcolor[rgb]{0.25,0.44,0.63}{##1}}}
\expandafter\def\csname PY@tok@ow\endcsname{\let\PY@bf=\textbf\def\PY@tc##1{\textcolor[rgb]{0.00,0.44,0.13}{##1}}}
\expandafter\def\csname PY@tok@sx\endcsname{\def\PY@tc##1{\textcolor[rgb]{0.78,0.36,0.04}{##1}}}
\expandafter\def\csname PY@tok@bp\endcsname{\def\PY@tc##1{\textcolor[rgb]{0.00,0.44,0.13}{##1}}}
\expandafter\def\csname PY@tok@c1\endcsname{\let\PY@it=\textit\def\PY@tc##1{\textcolor[rgb]{0.25,0.50,0.56}{##1}}}
\expandafter\def\csname PY@tok@kc\endcsname{\let\PY@bf=\textbf\def\PY@tc##1{\textcolor[rgb]{0.00,0.44,0.13}{##1}}}
\expandafter\def\csname PY@tok@c\endcsname{\let\PY@it=\textit\def\PY@tc##1{\textcolor[rgb]{0.25,0.50,0.56}{##1}}}
\expandafter\def\csname PY@tok@mf\endcsname{\def\PY@tc##1{\textcolor[rgb]{0.13,0.50,0.31}{##1}}}
\expandafter\def\csname PY@tok@err\endcsname{\def\PY@bc##1{\setlength{\fboxsep}{0pt}\fcolorbox[rgb]{1.00,0.00,0.00}{1,1,1}{\strut ##1}}}
\expandafter\def\csname PY@tok@kd\endcsname{\let\PY@bf=\textbf\def\PY@tc##1{\textcolor[rgb]{0.00,0.44,0.13}{##1}}}
\expandafter\def\csname PY@tok@ss\endcsname{\def\PY@tc##1{\textcolor[rgb]{0.32,0.47,0.09}{##1}}}
\expandafter\def\csname PY@tok@sr\endcsname{\def\PY@tc##1{\textcolor[rgb]{0.14,0.33,0.53}{##1}}}
\expandafter\def\csname PY@tok@mo\endcsname{\def\PY@tc##1{\textcolor[rgb]{0.13,0.50,0.31}{##1}}}
\expandafter\def\csname PY@tok@mi\endcsname{\def\PY@tc##1{\textcolor[rgb]{0.13,0.50,0.31}{##1}}}
\expandafter\def\csname PY@tok@kn\endcsname{\let\PY@bf=\textbf\def\PY@tc##1{\textcolor[rgb]{0.00,0.44,0.13}{##1}}}
\expandafter\def\csname PY@tok@o\endcsname{\def\PY@tc##1{\textcolor[rgb]{0.40,0.40,0.40}{##1}}}
\expandafter\def\csname PY@tok@kr\endcsname{\let\PY@bf=\textbf\def\PY@tc##1{\textcolor[rgb]{0.00,0.44,0.13}{##1}}}
\expandafter\def\csname PY@tok@s\endcsname{\def\PY@tc##1{\textcolor[rgb]{0.25,0.44,0.63}{##1}}}
\expandafter\def\csname PY@tok@kp\endcsname{\def\PY@tc##1{\textcolor[rgb]{0.00,0.44,0.13}{##1}}}
\expandafter\def\csname PY@tok@w\endcsname{\def\PY@tc##1{\textcolor[rgb]{0.73,0.73,0.73}{##1}}}
\expandafter\def\csname PY@tok@kt\endcsname{\def\PY@tc##1{\textcolor[rgb]{0.56,0.13,0.00}{##1}}}
\expandafter\def\csname PY@tok@sc\endcsname{\def\PY@tc##1{\textcolor[rgb]{0.25,0.44,0.63}{##1}}}
\expandafter\def\csname PY@tok@sb\endcsname{\def\PY@tc##1{\textcolor[rgb]{0.25,0.44,0.63}{##1}}}
\expandafter\def\csname PY@tok@k\endcsname{\let\PY@bf=\textbf\def\PY@tc##1{\textcolor[rgb]{0.00,0.44,0.13}{##1}}}
\expandafter\def\csname PY@tok@se\endcsname{\let\PY@bf=\textbf\def\PY@tc##1{\textcolor[rgb]{0.25,0.44,0.63}{##1}}}
\expandafter\def\csname PY@tok@sd\endcsname{\let\PY@it=\textit\def\PY@tc##1{\textcolor[rgb]{0.25,0.44,0.63}{##1}}}

\makeatother

\providecommand*{\DUrole}[2]{%
  \ifcsname DUrole#1\endcsname%
    \csname DUrole#1\endcsname{#2}%
  \else
    \ifcsname docutilsrole#1\endcsname%
      \csname docutilsrole#1\endcsname{#2}%
    \else%
      #2%
    \fi%
  \fi%
}

\ifthenelse{\isundefined{\hypersetup}}{
  \usepackage[colorlinks=true,linkcolor=blue,urlcolor=blue]{hyperref}
  \urlstyle{same} 
}{}

\begin{document}
\newcounter{footnotecounter}\title{Numerical simulation of liver perfusion: from CT scans to FE model}\author{Vladimír Lukeš$^{\setcounter{footnotecounter}{1}\fnsymbol{footnotecounter}\setcounter{footnotecounter}{2}\fnsymbol{footnotecounter}}$%
          \setcounter{footnotecounter}{1}\thanks{\fnsymbol{footnotecounter} %
          Corresponding author: \protect\href{mailto:vlukes@ntis.zcu.cz}{vlukes@ntis.zcu.cz}}\setcounter{footnotecounter}{2}\thanks{\fnsymbol{footnotecounter} University of West Bohemia, Pilsen, Czech republic}, Miroslav Jiřík$^{\setcounter{footnotecounter}{2}\fnsymbol{footnotecounter}}$, Alena Jonášová$^{\setcounter{footnotecounter}{2}\fnsymbol{footnotecounter}}$, Eduard Rohan$^{\setcounter{footnotecounter}{2}\fnsymbol{footnotecounter}}$, Ondřej Bublík$^{\setcounter{footnotecounter}{2}\fnsymbol{footnotecounter}}$, Robert Cimrman$^{\setcounter{footnotecounter}{2}\fnsymbol{footnotecounter}}$\thanks{%

          \noindent%
          Copyright\,\copyright\,2014 Vladimír Lukeš et al. This is an open-access article distributed under the terms of the Creative Commons Attribution License, which permits unrestricted use, distribution, and reproduction in any medium, provided the original author and source are credited. http://creativecommons.org/licenses/by/3.0/%
        }}\maketitle
          \renewcommand{\leftmark}{PROC. OF THE 7th EUR. CONF. ON PYTHON IN SCIENCE (EUROSCIPY 2014)}
          \renewcommand{\rightmark}{NUMERICAL SIMULATION OF LIVER PERFUSION: FROM CT SCANS TO FE MODEL}

\setcounter{page}{79}
\newcommand*{\docutilsroleref}{\ref}
\newcommand*{\docutilsrolelabel}{\label}
\AtEndDocument{\cleardoublepage}
\begin{abstract}We use a collection of Python programs for numerical simulation of
liver perfusion. We have an application for semi-automatic generation
of a finite element mesh of the human liver from computed tomography
scans and for reconstruction of the liver vascular structure. When the
real vascular trees can not be obtained from the CT data we generate
artificial trees using the constructive optimization method. The
generated FE mesh and vascular trees are imported into SfePy (Simple
Finite Elements in Python) and numerical simulations are performed in
order to get the pressure distribution and perfusion flows in the
liver tissue. In the post-processing steps we calculate transport of a
contrast fluid through the liver parenchyma.\end{abstract}\begin{IEEEkeywords}segmentation, liver perfusion, multicompartment model, finite element
method, finite volume method\end{IEEEkeywords}

\section{Introduction%
  \label{introduction}%
}

A patient specific numerical modelling of liver perfusion requires
cooperation of people with different specializations. Our research
group consists of cyberneticians, informaticians, mechanicians and
physicians who formulate problems to be solved and who are able to
judge the outcomes of mathematical models and simulations from the
point of view of medicine. The research is motivated by the needs of
surgeons, they would like to have efficient tools for better planning
of liver surgeries and who would also appreciate to be able to predict
changes of liver perfusion caused by diseases or after surgical
resections.

The task of numerical modelling of a human liver can be divided into
two sub-problems. First of all, the geometry of larger vascular
structures and hepatic parenchyma must be identified from data
obtained by computed tomography (CT) or magnetic resonance imaging
(MRI) examinations. With the knowledge of liver shape and vascular
structures, the numerical simulations of liver perfusion can be
performed using different mathematical models of blood flow at
different spatial scales. The question, how to obtain all the
necessary parameters of our models (permeabilities, etc.) is out of
the scope of this paper. More information can be found in \cite{Roh12b} or
\cite{Coo12}.

The mathematical model of tissue perfusion presented in this paper has
been already used for similar problems, e.g. for cardiac perfusion
\cite{Mic13}. With this in mind, our modelling approach seems to be
reasonable in the context of current computational biomechanics.

The main part of computation is performed in SfePy, which is an open
source Python framework for solving various problems described by
partial differential equations, see \cite{Cim14}, \cite{Cim14b}. Because we
participate in the development of the code, we have a full control
over the computational process and we are able to easily modify the
code or extend it to solve specific tasks. To extract geometrical data
of the liver and vascular structures from CT/MRI scans, we had to
develop own tools (DICOM2FEM, LISA, VTreeGen) due to unavailability of
any suitable open source applications that would meet our demands on
reliability, efficiency, portability and simplicity of use. The
processing toolchain for the numerical simulation of liver perfusion
is depicted in Fig. \DUrole{ref}{swtools}.\begin{figure}[]\noindent\makebox[\columnwidth][c]{\includegraphics[width=\columnwidth]{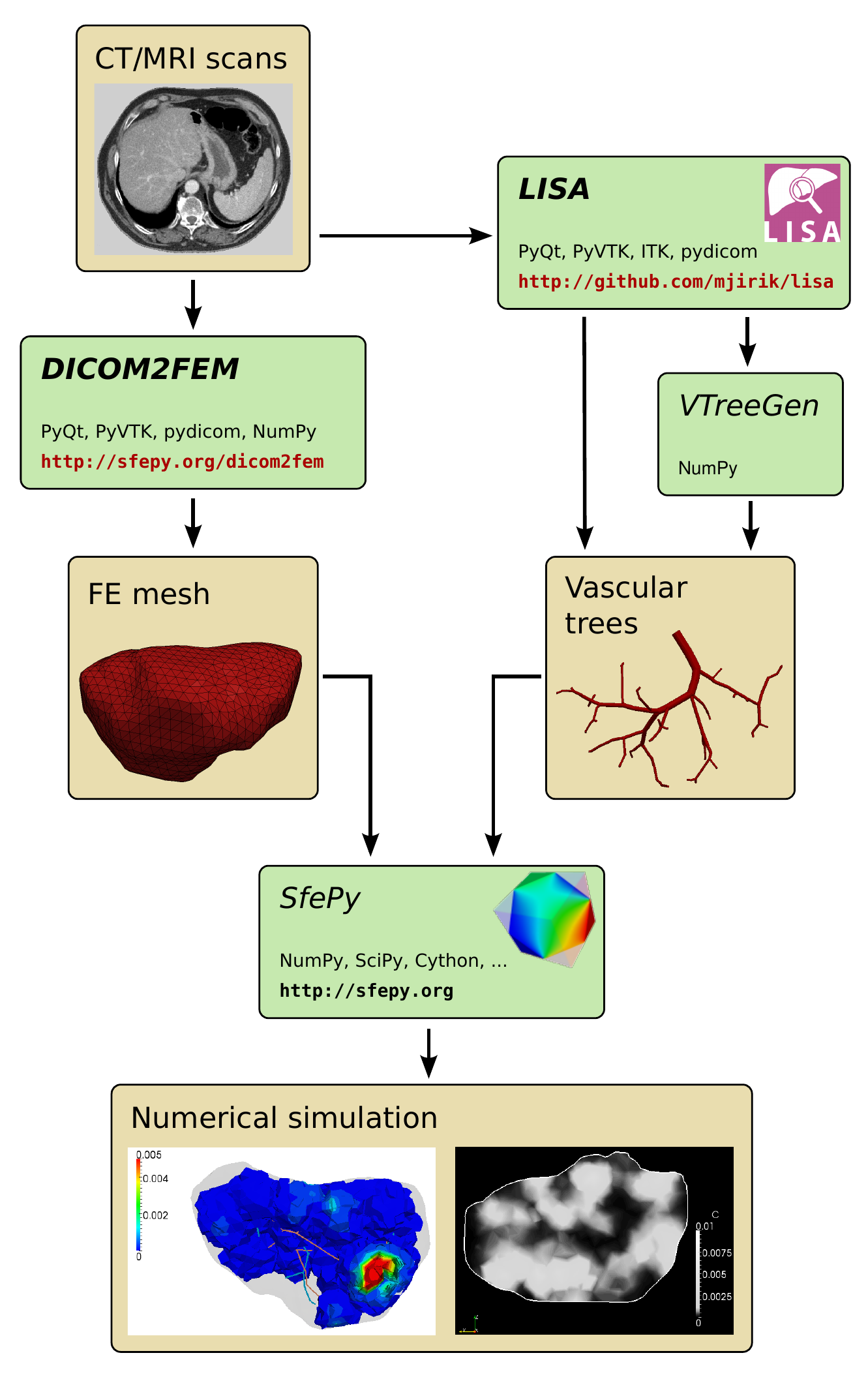}}
\caption{Toolchain for a patient specific numerical modelling of liver
perfusion. \DUrole{label}{swtools}}
\end{figure}

\section{Volumetric model of liver parenchyma%
  \label{volumetric-model-of-liver-parenchyma}%
}

We developed an application called DICOM2FEM \cite{Luk14} for semi-automatic
segmentation and generation of finite element meshes from CT scans
stored in the DICOM file format. Because it is a standard format for
storing information in medical imaging we are able to process data
from a wide range of sources (Magnetic Resonance Imaging, Positron
Emission Tomography, etc.). A series of the DICOM files is handled by
the pydicom library \cite{Mas14}. The graphical user interface of DICOM2FEM
(Fig. \DUrole{ref}{dicom2fem}) is build up using PyQt library and the
resulting data structures are stored to the VTK (Visualization
Toolkit) file by the help of PyVTK. Segmented bodies can be visualized
using a simple VTK viewer implemented in the application.

\subsection{Segmentation%
  \label{segmentation}%
}

Segmentation of the liver parenchyma from computed tomography or
magnetic resonance data is hard to solve because of low density
(intensity) contrast to adjacent organs like stomach or
hearth. Moreover, large individual anatomy differences should be taken
into account. Various methods for segmentation are compared in
\cite{Mha12} and \cite{Hei09}.

Our segmentation approach is based on the Graph-Cut method described
in \cite{Boy01} and \cite{Boy06}. We use the original implementation of
max-flow/min-cut algorithm \cite{Kol14} and the Python wrapper by Andreas
Müller \cite{Mül14}. Known weakness of the algorithm is great memory demand,
memory usage is quickly increasing as the data size grows. In our case
we define a region of interest and downsampling to suppress this
disadvantage. The Graph-Cut method combines advantages of region and
edge based segmentation methods. It minimizes cost function
$E(A)$:\begin{equation}
\label{segment1}
E(A) = \lambda \cdot R(A) + B (A),
\end{equation}where $A$ is the image segmentation, $R(A)$ is connected
to region properties of the image and $B(A)$ comprises boundary
properties of the segmentation. The coefficient $\lambda \geq 0$
specifies weight of the region $R(A)$ and the boundary term
$B(A)$.

For our purpose, the main benefit of this algorithm is a precise
control of the segmentation process. As it is shown in
Fig. \DUrole{ref}{dicom2fem}, the user (experienced in human anatomy) interactively
selects the liver tissue with the left mouse button (green seeds) and
the regions out of the liver with the right mouse button (red
seeds). Based on the seeds the density three component Gaussian
mixture model is estimated for the liver and the outer region. Using
the Gaussian model a graph representing input data and seeds is
constructed. By minimizing the cost function using the
max-flow/min-cut algorithm, the segmentation of the CT scans is
computed.\begin{figure}[]\noindent\makebox[\columnwidth][c]{\includegraphics[width=\columnwidth]{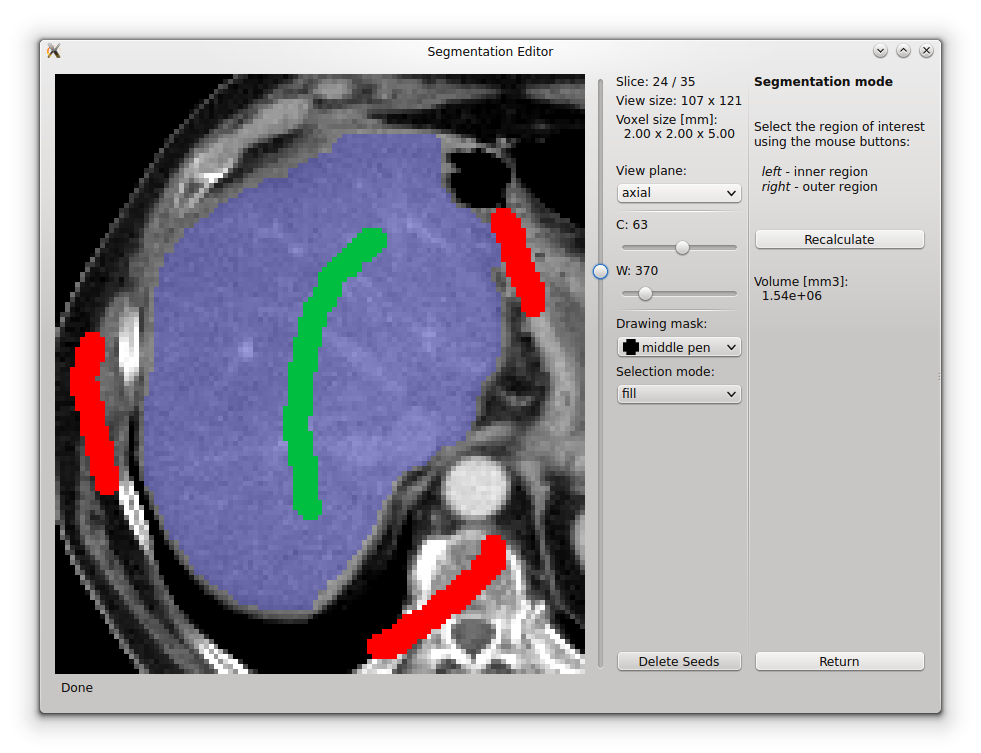}}
\caption{Segmentation editor for semi-automated segmentation of CT scans;
region of interest marked by green, region out of interest marked
by red. The selection should be made by a user experienced in the
anatomy. \DUrole{label}{dicom2fem}}
\end{figure}

\subsection{Mesh generation%
  \label{mesh-generation}%
}

The result of the segmentation process is a 3-dimensional binary array
(voxel array) together with information about the real size of the
voxels. The marching cubes algorithm \cite{Lor87} is used to generate
polygonal mesh of the organ surface. To improve the quality of surface
mesh, we apply the Taubin smoothing procedure \cite{Tau95} that is able to
preserve the total volume of the segmented organ. The smoothing
approach is based on signal processing on meshes, see Ref. \cite{Tau00},
and provides meshes of good quality.

The smoothed surface mesh is consequently processed by a tetrahedral
meshing function to get the volumetric FE model of the organ. The
marching cubes algorithm is computationally expensive so we implemented
the fast mesh generator (volumetric or surface), but it produces a
mesh with stair-step surface which can not be easily smoothed. The
fast generator is mainly used for testing purposes or for quick
preview of the FE model of body parts, for comparison of both
approaches see Fig. \DUrole{ref}{genmesh}.\begin{figure}[]\noindent\makebox[\columnwidth][c]{\includegraphics[width=\columnwidth]{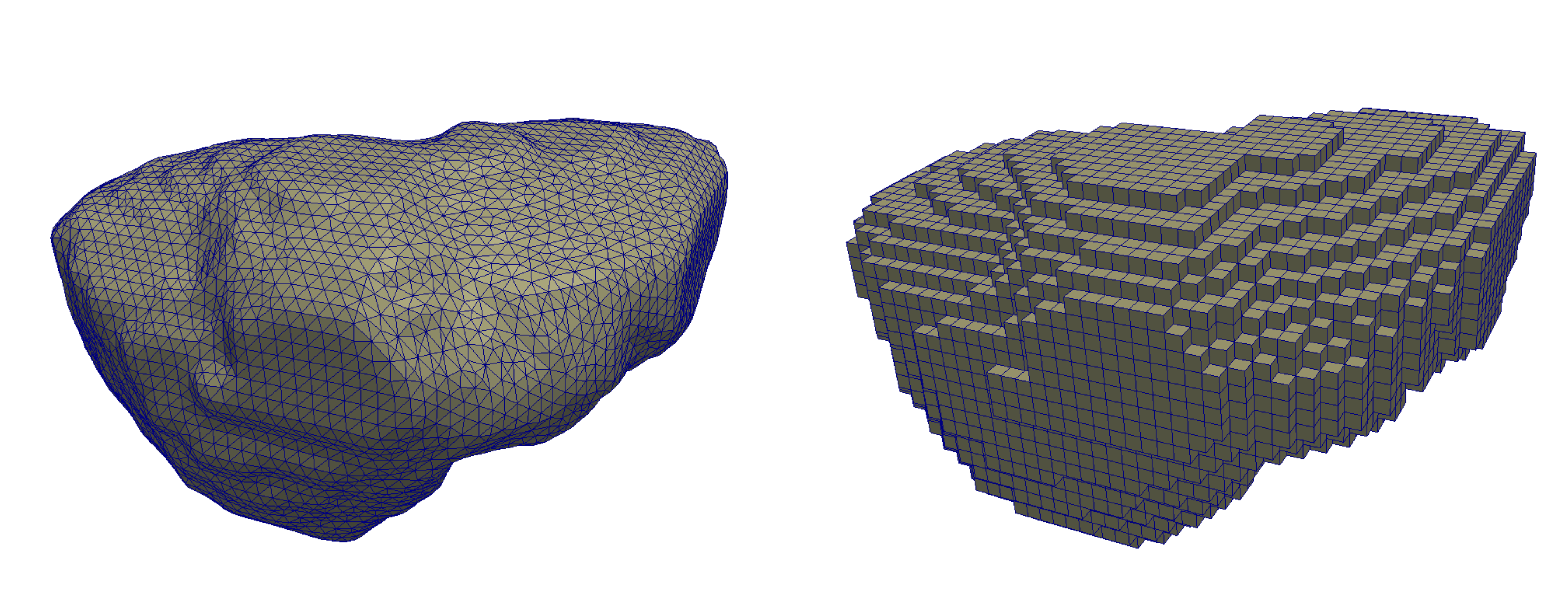}}
\caption{Finite element mesh of the liver parenchyma generated using: left)
marching cubes algorithm in combination with Taubin smoothing;
right) \textquotedbl{}voxel based\textquotedbl{} generator. \DUrole{label}{genmesh}}
\end{figure}

\section{Geometric model of vascular structures%
  \label{geometric-model-of-vascular-structures}%
}

1D models (trees) of the portal and hepatic vessels are crucial for
the numerical modelling of liver perfusion. We use them in computation
of fluxes and pressures in tree branches and in calculation of
transport times of contrast fluid within the vascular trees.

\subsection{Reconstruction of vascular structures%
  \label{reconstruction-of-vascular-structures}%
}

We obtain real vascular trees from CT scans using LISA (LIver Surgery
Analyser) \cite{Jir14}. It was developed as a tool for surgeons to help them
in a preoperative planning of liver resections. To be able to analyze
and detect the vascular structure, we need data form perfusion CT
examinations, when a contrast fluid is injected into the blood system
of a patient and the CT examinations are synchronized to capture the
filling period of the portal and hepatic systems in the liver. Due to
physiological conditions, automated detection of the liver portal tree
is an easier task then in the case of the hepatic tree.

Segmentation of a vessel tree is based on the algorithm described in
\cite{Sel02} with several modifications. In order to improve the quality
of CT data, we use the Gaussian blur denoising filter during the
preprocessing steps. We have automatic threshold selection based on a
histogram of the image, but the user is able to control this operation
by setting seed points. User interactivity is essential when
segmenting the vena cava, where the blood with a dissolved contrast
fluid is mixed with the blood from the rest of the body. The segmented
3D data are smoothed using a set of morphological operations - opening
and closing.\begin{figure}[]\noindent\makebox[\columnwidth][c]{\includegraphics[scale=0.30]{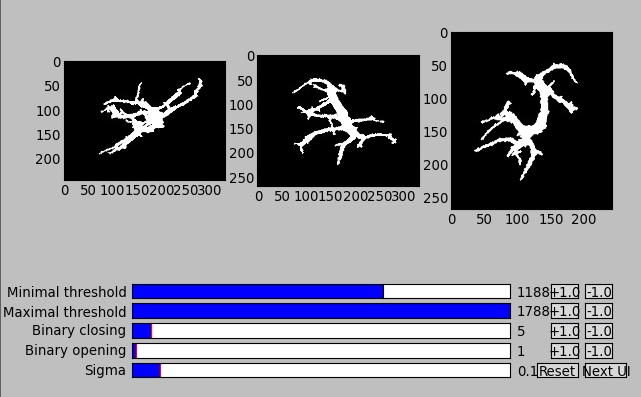}}
\caption{LISA - LIver Surgery Analyser: manual threshold selection for
vessel segmentation. \DUrole{label}{vesselseg}}
\end{figure}

A voxel-based representation of the vascular structures is transformed
into a graph representation preserving all important geometric
information (lengths, diameters, orientations, etc.) of the vessel
trees. ITK implementation of the 3D thinning algorithm \cite{Hom07} is
employed.



\subsection{Generation of artificial vascular trees%
  \label{generation-of-artificial-vascular-trees}%
}

Unfortunately, the detection of vascular structures is a very
complicated task with uncertain results. The input data are often
fuzzy or the vessels are not completely filled by the contrast fluid,
so the automated algorithm generates disconnected trees or trees with
various non-physiological artifacts. To avoid problems in further
simulation steps, we propose to take just a part of the reconstructed
tree and to generate the rest artificially using the constructive
optimization method \cite{Geo10}. This method is based on minimization of
intravascular blood volume and energy lost to friction. For global
optimization, a multilevel strategy with topological changes is used.

The whole optimization process consists of several steps: smoothing,
pruning and reconnecting. The smoothing step includes relaxation of
branching nodes which leads to a local minimum of the cost function in
a neighbourhood of a given node. When the relaxation places two
neighbour points at the same location, they are joined
together. During the smoothing operation, the branching points are
tested for splitting, this operation splits single branching into two
smaller to minimize the global cost. The splitting operation is
crucial step in the optimization process but also very computationally
expensive. In \cite{Geo10}, an efficient algorithm reducing this complexity
is proposed. The smoothing loop is repeated until the global cost is
minimized and further minimization can be achieved only by changing
the tree topology. Branches in a certain hierarchy, see
Fig. \DUrole{ref}{gentreehier}, (based on the Horton-Strahler order) are
discarded (pruned) and terminal nodes are reconnected to the nearest
nodes in the pruned tree. This procedure increases the freedom for the
optimization process and allows to find the better minimum of a cost
function. The smoothing-pruning-reconnecting loop is repeated several
times according to the number of hierarchies in the tree.\begin{figure}[]\noindent\makebox[\columnwidth][c]{\includegraphics[scale=0.20]{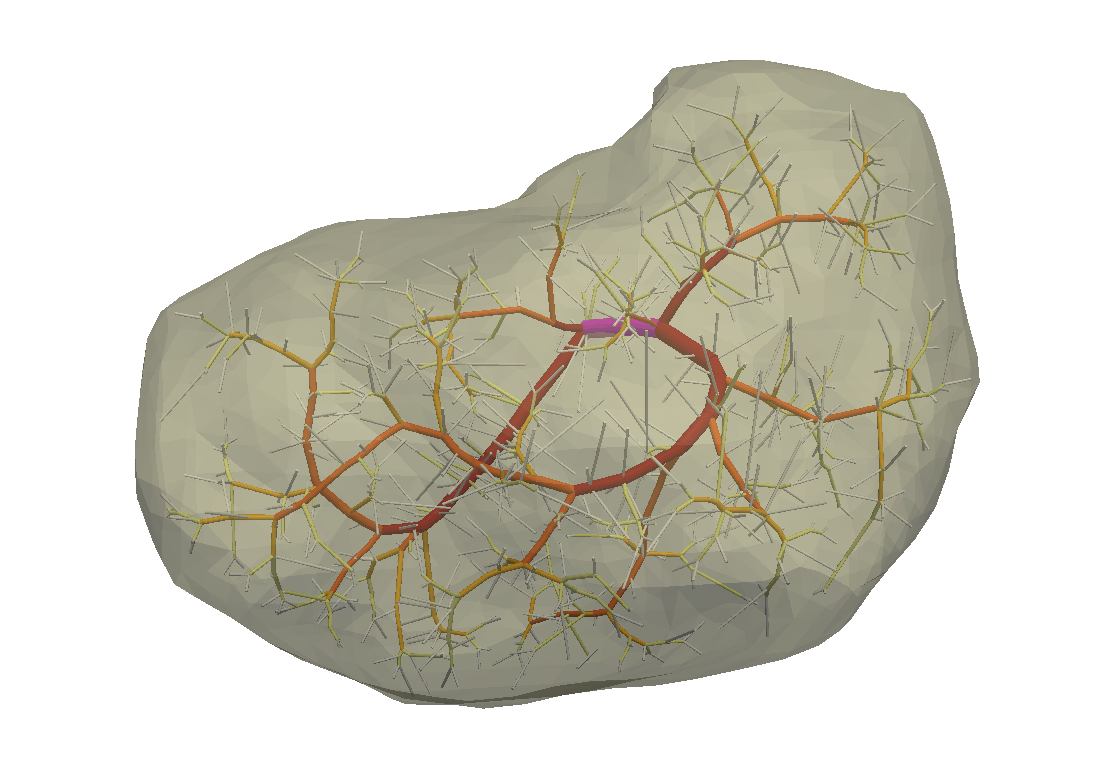}}
\caption{Hierarchy levels (distinguished by the colors) of the generated
portal tree, hierarchies based on the Horton-Strahler
order. \DUrole{label}{gentreehier}}
\end{figure}

We take the main branching part up to a certain diameter of the
vessels and generate randomly hundreds or thousands points inside the
liver volume. These points are considered as terminal points of the
vascular tree and are connected to the nearest branching points of the
reconstructed part. After the optimization, the artificial trees based
on real data with well defined hierarchy are obtained
(Fig. \DUrole{ref}{gentree}).\begin{figure}[]\noindent\makebox[\columnwidth][c]{\includegraphics[width=\columnwidth]{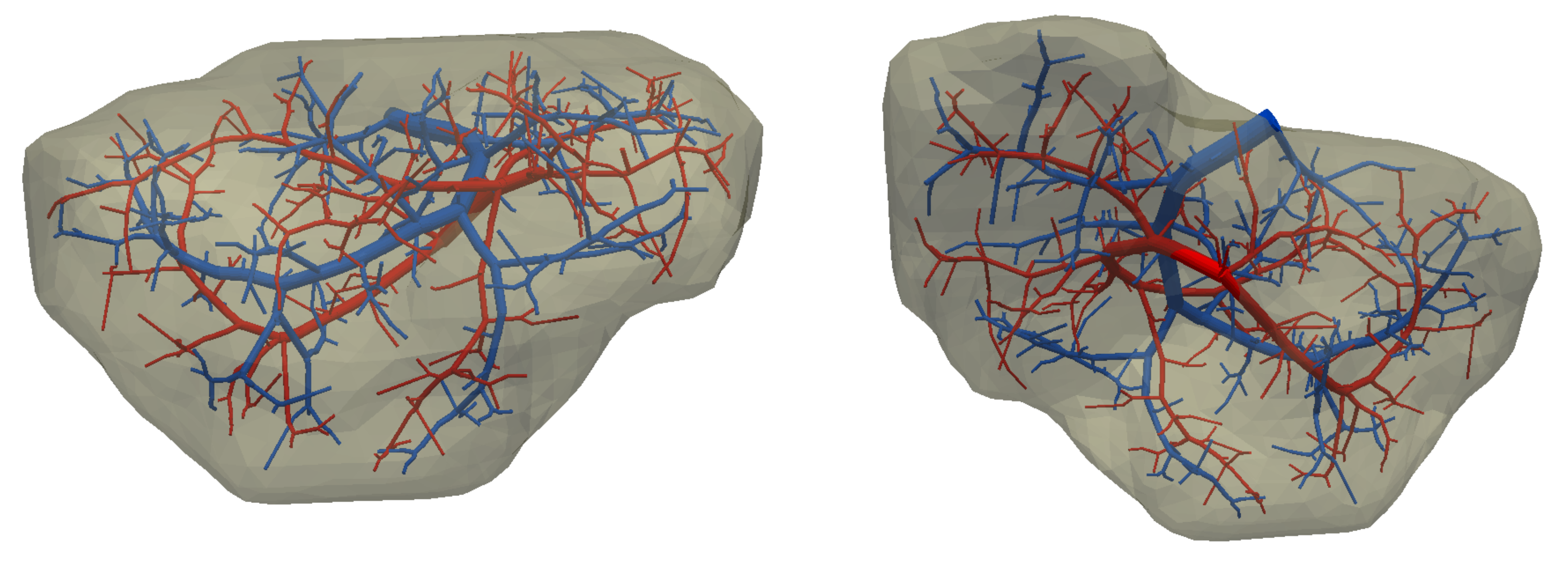}}
\caption{Generated vascular trees inside the liver volume representing
portal (red) and hepatic (blue) veins, front and rear
view. \DUrole{label}{gentree}}
\end{figure}

Our Python code for generating artificial vascular trees is called
VTreeGen, see Fig. \DUrole{ref}{swtools}. It is not yet publicly available as
other tools presented in this paper.

\section{Mathematical model of liver perfusion%
  \label{mathematical-model-of-liver-perfusion}%
}

Numerical modelling of blood flow through the human liver presents a
complex task due to a wide range of scales involved in the problem
making it necessary to use different mathematical models for each
group of scales. The flow in branching vessels with diameters above 2
mm is described by a simple 1D model based on the Bernoulli equation
while the blood flow at lower hierarchies is modelled as parallel
flows in a 3D porous media governed by the Darcy's equation. Spatially
co-existing domains are referred as compartments, each of them
reflects a certain hierarchy of the tissue vascularity. The
compartments are coupled together and communicate with the 1D flow
model through sources and sinks, see Refs. \cite{Roh12}, \cite{Roh12b},
\cite{Mic13}, \cite{Joa14}.

The multicompartment approach allows to respect the different
characteristic features of perfusion hierarchies present in the tissue
parenchyma. Each compartment is associated with a permeability tensor
that reflects the vascular structure (vessel size and orientation) at
a given hierarchy level. The fluid exchange between different
compartments is driven by a coupling coefficient.

\subsection{Blood flow in vascular trees%
  \label{blood-flow-in-vascular-trees}%
}

We assume that the simple 1D flow model gives sufficient accuracy in
the context of our simulations. The main advantage of the 1D model
is the minimal computational cost compared to a full 3D flow
simulation which obviously would give more realistic results. A
detailed study of 3D and 1D flow models can be found in \cite{Joa14b}.

The mathematical model of the flow in the branching tree can be
described by the mass conservation and Bernoulli equations:\begin{equation}
\label{bernoulli1}
A_0 w_0 = \sum_k^{n}   A_k w_k,
\end{equation}\begin{equation}
\label{bernoulli2}
\frac{1}{2}\rho w_0^2 + p_0 = \frac{1}{2}\rho w_k^2 +
p_k+e_k^\textrm{loss}\;,\quad k = 1,2,\dots,n,
\end{equation}where $A_k$ is the cross-section of branches and $n$ is
the number of terminal nodes (sources/sinks) connected to the liver
parenchyma.

The terms $e_k^\textrm{loss}$ represent the friction loss in
inelastic tubes and are defined as:\begin{equation}
\label{bernoulli3}
e_k^\textrm{loss}=\frac{1}{2}\varrho w_k^2
\frac{L}{D}\frac{64}{\textrm{Re}_k},
\end{equation}where $\varrho$ is the fluid density, $L$ and $D$
are the length and diameter of the branch and ${\textrm{Re}_k}$
is the Reynolds number.

The system of non-linear algebraic equations (\DUrole{ref}{bernoulli1}) and
(\DUrole{ref}{bernoulli2}) can be solved numerically using the Newton
method. For a given input velocity $w_0$ and terminal pressures
$p_k$, we are able to calculate the unknown input pressure
$p_0$ and terminal velocities $w_k$.

\subsection{Darcy flow in parenchyma%
  \label{darcy-flow-in-parenchyma}%
}

We assume a simple idealized model of liver perfusion comprising of
three co-existing compartments. The first one is attached to the 1D
portal venous tree such that the terminal branches of the tree are
local sources for the Darcy model. The second (middle) compartment
represents the filtration system of liver lobules. The last
compartment is connected to the hepatic vessel tree and the connecting
points play a role of sinks of the perfusion system, see
Fig. \DUrole{ref}{compartments}\begin{figure}[]\noindent\makebox[\columnwidth][c]{\includegraphics[scale=0.30]{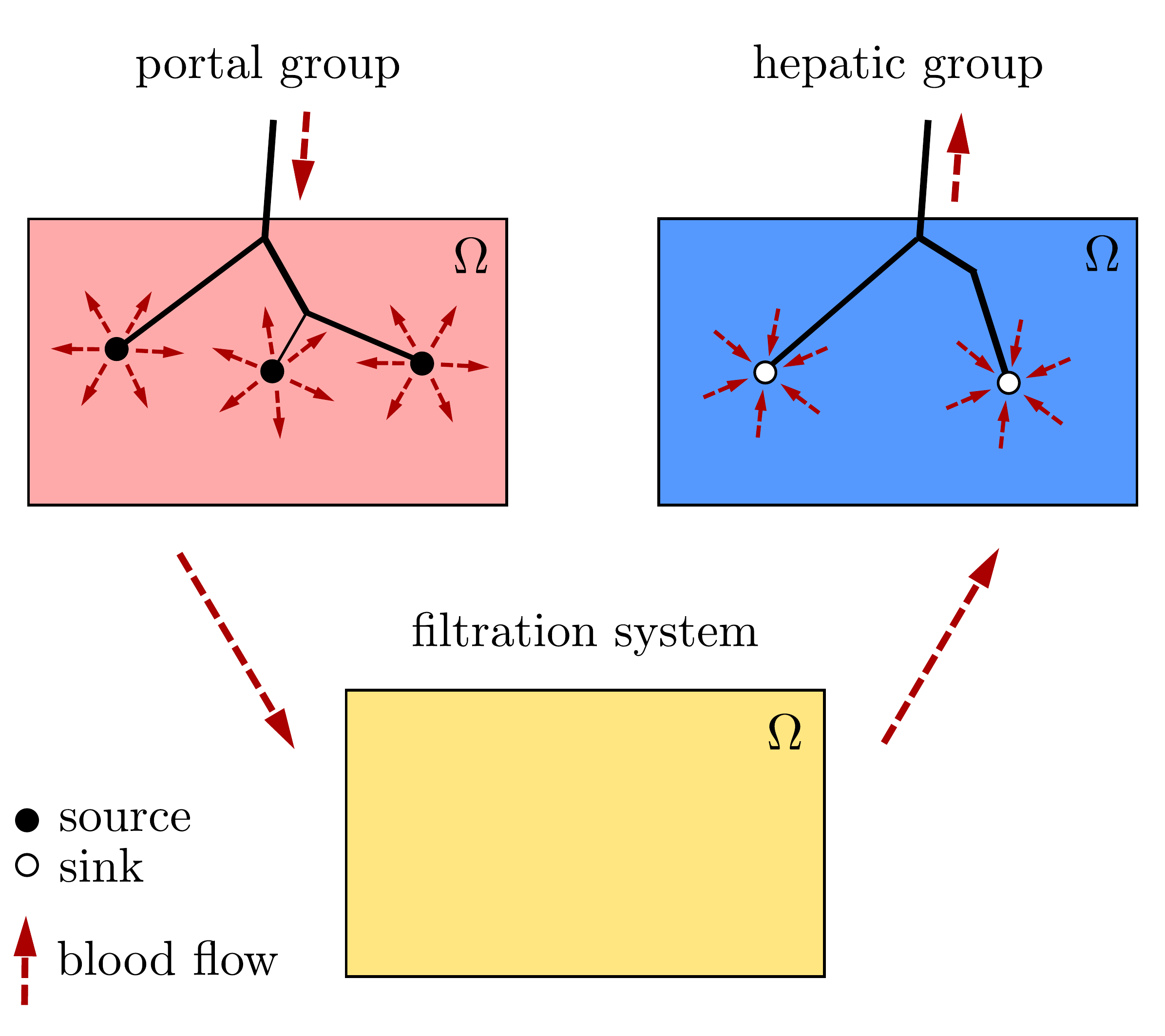}}
\caption{Schematic drawing of spatially co-existing compartments
(representing distinct perfusion hierarchies in the liver
parenchyma) connected to the reconstructed or generated portal and
hepatic venous trees via sources and sinks. \DUrole{label}{compartments}}
\end{figure}

The multicompartment Darcy system of $N$ compartments can be
written as:\begin{equation}
\label{darcy1}
\nabla \cdot \mathbf{w}^i + \sum_{j} G_j^i(p^i - p^j) = f^i,
\end{equation}\begin{equation}
\label{darcy2}
\mathbf{w}^i = - \mathbf{K}^i \nabla p^i,
\end{equation}for $i = 1\dots N$, where $\mathbf{K}^i$ is the local
permeability of the $i$-th compartment network and $G_j^i$
is the perfusion coefficient related to compartments $i$,
$j$, so that $G_j^i(p^i - p^j)$ describes the amount of
fluid going from $i$ to $j$ ($G_j^i$ is symmetric,
i.e. $G_j^i$ = $G_i^j$). In our case, when only three
compartments are considered, as shown in Fig. \DUrole{ref}{compartments}, we
take coupling parameters $G_1^2$, $G_2^3$ (and also
$G_2^1$, $G_3^2$) $\neq 0$, otherwise $G_i^j =
0$.

The discretized perfusion model is based on the weak formulation of
(\DUrole{ref}{darcy1}) and (\DUrole{ref}{darcy2}): Find $p_i \in V^i$ such
that for all $q_i \in V^i_0$:\begin{equation}
\label{darcy3}
 \int_{\Omega} \mathbf{K}^i \nabla p^i \cdot \nabla q^i +
 \int_{\Omega} \sum_j G_{j}^i(p^i - p^j) q^i = \int_{\Omega} f^i
 q^i,
\end{equation}for all compartments $i = 1,\dots,N$, where $V^i$,
$V^i_0$ are admissible sets, for more details see
Ref. \cite{Roh12}.

The multicompartment Darcy flow model is implemented in SfePy (Simple
Finite Elements in Python), see \cite{Cim14}, \cite{Cim14b}. SfePy is a framework
for solving various kinds of problems (mechanics, physics, biology, ...)
described by partial differential equations in two or three
space dimensions by the finite element method. The code is written
mostly in Python (C and Cython are used in some places due to
speed). Solvers and algorithms from SciPy \cite{Jon14} are used as well.

\subsection{Transport of contrast fluid%
  \label{transport-of-contrast-fluid}%
}

To assess the computed liver perfusion and possibly to compare the
numerical results with real perfusion data in the future, a dynamic
perfusion test is simulated. This test involves the modelling of
contrast fluid (tracer) transport through the hepatic tissue using the
perfusion velocities computed by the Darcy flow model for each of the
parenchyma compartments. The equations governing not only the
transport of the contrast fluid within one compartment, but also its
exchange between several compartments are numerically solved using an
upwind cell-centered finite volume scheme formulated for unstructured
grids in combination with the second-order accurate two-stage
Runge-Kutta method \cite{Joa14}.

\subsection{Numerical results%
  \label{numerical-results}%
}

The results of numerical simulations of tissue perfusion in the three
compartment model are shown in Fig. \DUrole{ref}{simulation1}, where the
computed perfusion velocities in the filtration (inter) and hepatic
compartments are depicted.

For illustration, Fig. \DUrole{ref}{simulation2} shows the tracer
distribution in an image-based model of human liver at selected time
instants including the corresponding total concentration $C$ as
would be seen in a CT scan. Here, the content of the tracer dissolved
in the blood in each compartment (portal, filtration and hepatic) is
expressed by the saturation $S$.\begin{figure}[]\noindent\makebox[\columnwidth][c]{\includegraphics[width=\columnwidth]{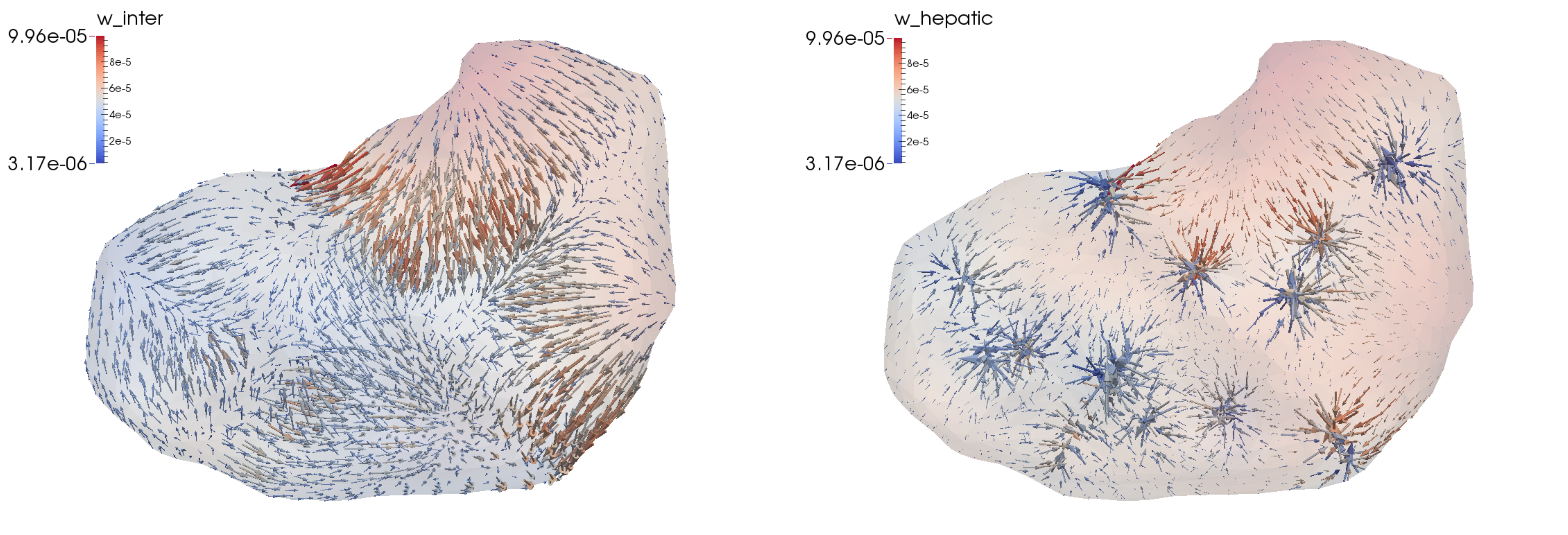}}
\caption{Computed perfusion velocities in the filtration (left) and hepatic (right)
compartments are depicted. \DUrole{label}{simulation1}}
\end{figure}\begin{figure}[]\noindent\makebox[\columnwidth][c]{\includegraphics[width=\columnwidth]{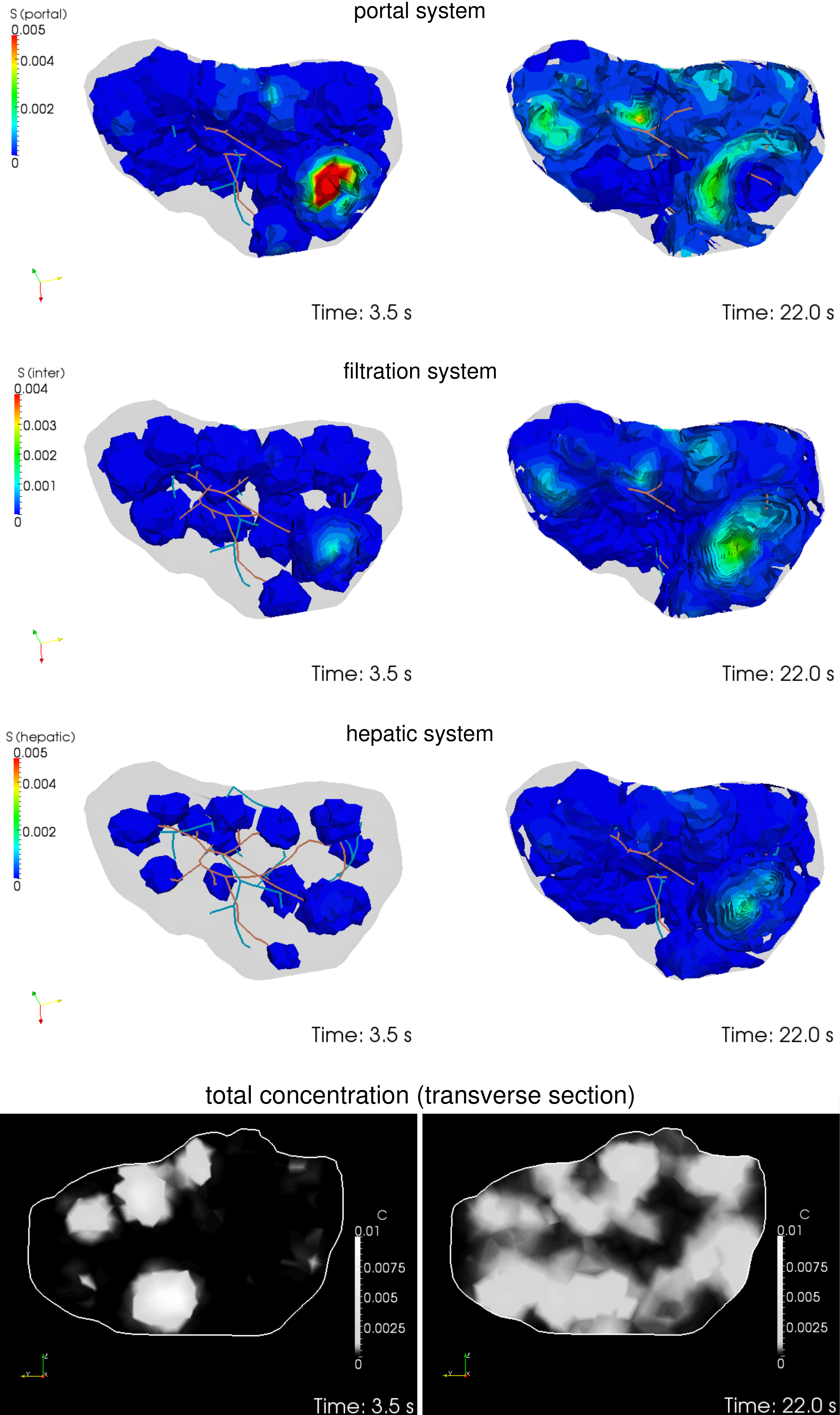}}
\caption{Time development of the saturation in the portal, filtration
(inter) and hepatic systems and the corresponding distribution of
the total concentration $C$ in the transverse
section. \DUrole{label}{simulation2}}
\end{figure}

\section{Conclusion%
  \label{conclusion}%
}

Using Python with standard modules for scientific computing and image
processing together with the Python based finite element solver SfePy
and a collection of developed supporting applications, we are able to
produce a simplified patient specific liver model and numerically
simulate hepatic blood perfusion.

CT data are processed by the semi-automatic segmentation algorithms
generating 1D structures representing the vascular trees and a 3D
volumetric model of the liver tissue. In case of incomplete or
unreliable results of the vascular trees reconstruction, we fabricate
artificial trees using constructive optimization approach. The
reconstructed or fabricated 1D trees and the volumetric liver model
are employed in numerical simulations of liver blood perfusion using
the finite element method. The model of contrast fluid propagation
provides time-dependent concentration of the tracer, that can be
compared with the standard medical measurements. It will allow us to
solve the inverse problem in order to identify some of the perfusion
parameters of our models. This is a crucial point for further
development.

Despite the fact that there is still a wide gap between our current
research and clinical practice, the LISA application was successfully
tested by radiologists and surgeons for volumetric analyses of livers
prior to surgeries and is now actively used.

\subsection{Acknowledgment%
  \label{acknowledgment}%
}

This research is partially supported by the Ministry of Health of the
Czech Republic, project NT 13326, and by the European Regional
Development Fund (ERDF), project \textquotedbl{}NTIS - New Technologies for the
Information Society\textquotedbl{}, European Centre of Excellence,
CZ.1.05/1.1.00/02.0090.


\begin{thebibliography}{Cim14b}
\bibitem[Boy01]{Boy01}{

Y. Boykov, O. Veksler, R. Zabih. \emph{Fast approximate energy
minimization via graph cuts.} In Pattern Analysis and
Machine Intelligence, 23(11):1222-1239, 2001.}
\bibitem[Boy06]{Boy06}{

Y. Boykov, G. Funka-Lea. \emph{Graph Cuts and Efficient N-D Image Segmentation.}
In International Journal of Computer Vision, 70:109–131, 2006.}
\bibitem[Cim14]{Cim14}{

R. Cimrman, et al. \emph{SfePy, finite element code and applications.}
Home page: \url{http://sfepy.org} {[}Accessed 2014-08-20{]}.}
\bibitem[Cim14b]{Cim14b}{

R. Cimrman. \emph{SfePy - Write Your Own \{FE\} Application.}
In Proceedings of the 6th European Conference on Python in
Science (EuroSciPy 2013), pages 65-70, 2014. \url{http://arxiv.org/abs/1404.6391}.}
\bibitem[Coo12]{Coo12}{

A. N. Cookson, J. Lee, C. Michler, R. Chabiniok, E. Hyde,
D. A. Nordsletten, M. Sinclair, M. Siebes, N. P. Smith.
\emph{A novel porous mechanical framework for modelling the
interaction between coronary perfusion and myocardial
mechanics.} In Journal of Biomechanics, 45(5):850-855, 2012.}
\bibitem[Geo10]{Geo10}{

M. Georg, T. Preusser, H. K. Hahn. \emph{Global Constructive
Optimization of Vascular Systems.} Technical Report:
Washington University in
St. Louis. \url{http://cse.wustl.edu/Research/Lists/TechnicalReports/Attachments/910/idealvessel_1.pdf}.}
\bibitem[Hei09]{Hei09}{

Heimann et al. \emph{Comparison and evaluation of methods for
liver segmentation from CT datasets.} In IEEE Transactions
on Medical Imaging, 28(8):1251-1265, 2009.}
\bibitem[Hom07]{Hom07}{

H. Homann. \emph{Implementation of a 3D thinning algorithm.} In
Insight Journal, July - December, 2007.}
\bibitem[Jir14]{Jir14}{

M. Jiřík. \emph{LISA - LIver Surgery Analyser.} Home page:
\url{https://github.com/mjirik/lisa} {[}Accessed 2014-08-20{]}.}
\bibitem[Joa14]{Joa14}{

A. Jonášová, E. Rohan, V. Lukeš, O. Bublík. \emph{Complex
hierarchical modeling of the dynamic perfusion test:
application to liver.} In Proceedings of 11th World Congres
of Computational Mechanics, 2014.}
\bibitem[Joa14b]{Joa14b}{

A. Jonášová, O. Bublík, E. Rohan, J. Vimmr. \emph{Simulation of
contrast medium propagation based on 1D and 3D portal
hemodynamics.} In: Proc. of the 20th International
Conference Engineering Mechanics, Svratka, Czech
Republic, 2014.}
\bibitem[Jon14]{Jon14}{

E. Jones, T. E. Oliphant, P. Peterson, et al. \emph{SciPy: Open
source scientific tools for Python.} Home page:
\url{http://www.scipy.org} {[}Accessed 2014-08-20{]}.}
\bibitem[Kol14]{Kol14}{

V. Kolmogorov. \emph{Max-flow/min-cut.} Home page:
\url{http://vision.csd.uwo.ca/code/} {[}Accessed 2014-08-20{]}.}
\bibitem[Lor87]{Lor87}{

W. E. Lorensen, H. E. Cline. \emph{Marching Cubes: A high
resolution 3D surface construction algorithm.} Computer
Graphics, Vol. 21, Nr. 4, 1987.}
\bibitem[Luk14]{Luk14}{

V. Lukeš. \emph{DICOM2FEM - application for semi-automatic
generation of finite element meshes.} Home page:
\url{http://sfepy.org/dicom2fem} {[}Accessed 2014-08-20{]}.}
\bibitem[Mas14]{Mas14}{

D. Mason. \emph{pydicom}, available at
\url{https://code.google.com/p/pydicom/} {[}Accessed 2014-08-20{]}.}
\bibitem[Mha12]{Mha12}{

A. M. Mharib, A. R. Ramli, S. Mashohor, R. B. Mahmood. \emph{Survey
on liver CT image segmentation methods.} In Artificial
Intelligence Review, 37(2):83-95, 2012.}
\bibitem[Mic13]{Mic13}{

C. Michler, A. Cookson, R. Chabiniok, E. Hyde, J. Lee,
M. Sinclair, T. Sochi, A. Goyal, G. Vigueras, D. Nordsletten, N. Smith.
\emph{A computationally efficient framework for the simulation
of cardiac perfusion using a multi-compartment Darcy
porous-media flow model.} Int. Journal for Numerical
Methods in Biomedical Engineering, 29(2):217-32, 2013.}
\bibitem[Mül14]{Mül14}{

A. Müller. \emph{Python wrappers for GCO alpha-expansion and
alpha-beta-swaps.} Home page:
\url{https://github.com/amueller/gco_python} {[}Accessed 2014-08-20{]}.}
\bibitem[Oli07]{Oli07}{

T. E. Oliphant. \emph{Python for scientific computing.} In
Computing in Science \& Engineering,
9(3):10-20, 2007. Home page: \url{http://www.numpy.org}.}
\bibitem[Tau95]{Tau95}{

G. Taubin. \emph{A signal processing approach to fair surface
design.} In Siggraph'95 Conference Proceedings, pages
351–358, 1995.}
\bibitem[Tau00]{Tau00}{

G. Taubin. \emph{Geometric Signal Processing on Polygonal
Meshes.}, In EUROGRAPHICS 2000, 2000.}
\bibitem[Roh12]{Roh12}{

E. Rohan, V. Lukeš, A. Jonášová, O. Bublík. \emph{Towards
microstructure based tissue perfusion reconstruction from
CT using multiscale modeling.} In Proc. of the 10th World
Congress on Computational Mechanics, Sao Paulo, Brasil, 2012.}
\bibitem[Roh12b]{Roh12b}{

E. Rohan, V. Lukeš. \emph{Modeling tissue perfusion using a
homogenized model with layer-wise decomposition.}
In Preprints MATHMOD 2012, Vienna University of Technology,
Austria, (2012).}
\bibitem[Sel02]{Sel02}{

D. Selle, B. Preim, A. Schenk, H. O. Peitgen. \emph{Analysis of
vasculature for liver surgical planning.} In IEEE
Transactions on Medical Imaging, 21(11):1344-1357, 2002.}
\end{thebibliography}
\end{document}